\documentclass[%
 reprint,
 amsmath,
 amssymb,
 aps,
]{revtex4-1}
\usepackage{subfigure}
\usepackage{graphicx}
\usepackage{dcolumn}
\usepackage{bm}

\begin{document}

\preprint{APS/123-QED}

\title{Radiation friction force effects on electron dynamics in ultra-intensity laser pulse}

\author{Yanzeng Zhang}
\author{Sergei Krasheninnikov}%
\affiliation{%
 Mechanical and Aerospace Engineering Department, University of California San Diego, La Jolla, CA 92093, USA
}%


\begin{abstract}
The electron dynamics in the ultra-high intensity laser pulse with radiation friction force in the Landau-Lifshitz form are studied. It is demonstrated that widely used approximation, where only the term dominating the dissipation of electron kinetic energy is retained in the expression for the radiation friction, is incorrect for the case of diverging electron trajectories. As a matter of fact, for large friction force effects, all components of the radiation friction force in the Landau-Lifshitz form have the same order in the equation of electron motion, being equally important for both electron trajectory and thus energy gain in the case of diverging electron trajectories (e.g. determined by the superposition of few electromagnetic waves).

\end{abstract}

\pacs{Valid PACS appear here}
\maketitle
It is well known that electron dynamics in a strongly relativistic electromagnetic field could be significantly altered by so called “radiation damping” effects \cite{11LDLandaubook,2YBZel}. In some contemporary literature these effects are also called the “radiation friction” effects. In particular, radiation friction becomes important for the case of plasma interactions with ultra-intense laser pulse (e.g. see \cite{3ESarachikGSch,4Zhidkov,9bulanov2011lorentz,7ARBell,9GLehmann,6SVBulanov,8JGKirk,10Agonoskov,11TZhEsirkepov,12SVBulanov2017,5LLjiAPukov} and the references therein). 

Recently it was found that for some cases the radiation friction results in electron trapping in the vicinity of some spatially localized points (attractors)  \cite{8JGKirk,10Agonoskov,11TZhEsirkepov,12SVBulanov2017,5LLjiAPukov}. However, these studies were performed by using just a part of the Landau-Lifshitz expression \cite{11LDLandaubook} for the radiation friction, which dominates in the dissipation of electron kinetic energy. We will see that this approximation is incorrect for the case of diverging electron trajectories corresponding to electron dynamics near such attractors.

In this study we consider an impact of the radiation friction on electron dynamics exposed to the combination of monochromatic waves with the same frequency but opposite propagation directions, which is widely used in theoretical studies (e.g. see \cite{6SVBulanov,7ARBell,8JGKirk,9GLehmann,10Agonoskov,11TZhEsirkepov,12SVBulanov2017}). We use standard normalized variables of $\hat{t}=t\omega$, $\hat{\mathbf{x}}=k\mathbf{x}$, $\hat{\mathbf{v}}=\mathbf{v}/c$, and $(\hat{\mathbf{E}},\hat{\mathbf{B}})=e(\mathbf{E},\mathbf{B})/m\omega c$, where $e$ is the elementary charge, $m$ is electron mass, $c$ is the light speed, and $\omega=kc$ and $k$ are, respectively, the wave frequency and wavenumber. Then the equation of electron motion in the electromagnetic field with radiation friction force in the Landau-Lifshitz form of
\begin{equation}
\mathbf{f}_{RF}=\rho_f(\gamma^2\mathbf{f}_1+\gamma\mathbf{f}_2+\mathbf{f}_3),
\label{eq1}
\end{equation}
where $\gamma$ is the relativistic factor, $\rho_{f}=2r_ek/3\ll 1$, $r_e=e^2/mc^2$ is the classical electron radius, and
\begin{widetext}
\begin{equation}
\mathbf{f}_1=-\mathbf{v}[(\mathbf{E}+\mathbf{v}\times \mathbf{B})^2-(\mathbf{v}\cdot \mathbf{E})^2],\mathbf{f}_2=[(\partial_t+\mathbf{v}\cdot\bigtriangledown)\mathbf{E}+ \mathbf{v}\times(\partial_t+\mathbf{v}\cdot\bigtriangledown)\mathbf{B}],\mathbf{f}_3=[ \mathbf{E}\times\mathbf{B}+\mathbf{B}\times(\mathbf{B}\times\mathbf{v})+\mathbf{E}(\mathbf{v}\cdot\mathbf{E}) ],
 \label{eq2}
\end{equation}
\end{widetext}
can be written as follows 
\begin{eqnarray}
& &\frac{d\mathbf{P}}{dt}=-(\mathbf{E}+\mathbf{v}\times\mathbf{B})+Q(q)\mathbf{f}_{RF}, \label{eq3}\\
& &\frac{d\gamma}{dt}=-\mathbf{E}\cdot \mathbf{v}+Q(q)\mathbf{f}_{RF}\cdot\mathbf{v}, \label{eq4}
\end{eqnarray}
where $\mathbf{P}=\gamma m \mathbf{v}$ and the function $Q(q)$ of
\begin{equation}
q=\frac{\gamma}{a_s}\sqrt{(\mathbf{E}+\mathbf{v}\times \mathbf{B})^2-(\mathbf{v}\cdot \mathbf{E})^2},
\label{eq5}
\end{equation}
characterizes the quantum effects with $a_s=mc/\hbar k$ being the normalized Schwinger field, and $\hbar$ the Planck constant (e.g. see \cite{6SVBulanov} and the references therein). In \cite{11TZhEsirkepov} it was shown that for $q\le10$, $Q(q)$ could be approximated as 
\begin{equation}
Q(q)\approx(1+18q+69q^2+73q^3+5.806q^4)^{-1/3}.
\label{eq6}
\end{equation}
We notice that in Eqs.~(\ref{eq2}-\ref{eq5}) we removed hats over the normalized variables to simplify the expressions. For the discussion of the applicability of the Landau-Lifshitz expression for the radiation friction see \cite{9bulanov2011lorentz} and the references therein. 

For modest quantum effects, $Q(q)\sim 1$ , and assuming that $\gamma\approx a_0$, where $a_0$ is the normalized amplitude of laser wave vector potential, from Eq.~(\ref{eq4}) we find that the contributions of different components of the friction force, $\mathbf{f}_i$ (i=1, 2, 3), to the Eq.~(\ref{eq4}) for electron kinetic energy could be estimated as follows: $f_1\sim a_0^4$ and $f_2,f_3\sim a_0^2$ (we use $f=\left|\mathbf{f}\right|$ to simplify the expression when estimating forces magnitude). As a result, in super-relativistic regime where $\gamma\approx a_0\gg 1$, $\mathbf{f}_1$ component of the friction force dominates in Eq.~(\ref{eq4}) and it starts to compete with the Lorentz force for $\eta_f=a_0^3\rho_f\ge 1$ . Based on this estimate, when investigating the impact of the radiation friction on electron dynamics in ultra-intense laser pulse where the Landau-Lifshitz form of radiation friction force was used, only $\mathbf{f}_1$  component is often taken into account. 

However, electron energy gain/loss depends also on the magnitude of laser wave, which is a function of the spatial coordinate. Therefore, it's necessary to evaluate an impact of the radiation friction on electron trajectory. From Eqs.~(\ref{eq1}-\ref{eq4}) we find 
\begin{eqnarray}
\gamma\frac{d\mathbf{v}}{dt}&=&-(\mathbf{E}+\mathbf{v}\times\mathbf{B})+(\mathbf{E}\cdot\mathbf{v})\mathbf{v} \label{eq7}\\
& &+Q(q)\rho_{f} \{\mathbf{f}_1+\gamma[\mathbf{f}_2-(\mathbf{f}_2\cdot\mathbf{v})\mathbf{v}]+[\mathbf{f}_3-(\mathbf{f}_3\cdot\mathbf{v})\mathbf{v}]\}\nonumber,
\end{eqnarray}
where the factor of $\gamma^2=1-v^2$ in front of $\mathbf{f}_1$ component is eliminated as $\mathbf{f}_1$ is aligned with electron velocity as seen from Eq.~(\ref{eq2}), but not for $\mathbf{f}_2$ and $\mathbf{f}_3$ components. As a result, one could expect that for $\gamma\approx a_0$,  the contributions of all components of the radiation friction to electron trajectory are comparable. This is in contrast to the equation for kinetic electron energy (\ref{eq4}), where $\mathbf{f}_1$ component dominates. 

We notice that Eqs.~(\ref{eq3}, \ref{eq4}, \ref{eq7}) neglect the force $\mathbf{f}_s$, related to the interaction of electron’s spin with electromagnetic field (e.g. see \cite{14MTamburini,15SMMahajan,16MWenHBauke} and the references therein), the magnitude of which compared with those of the radiation friction force could be estimated as \cite{9GLehmann,14MTamburini,15SMMahajan,16MWenHBauke}: $f_s/f_2\sim 1/\alpha \gamma$, $f_s/f_1\sim f_s/f_3\sim 1/\alpha a_0$, where $\alpha=e^2/\hbar c=1/137$ is the fine structure constant. In our simulations we choose such $a_0$ that $f_s<f_1,f_3$ and thus we neglect the force $\mathbf{f}_s$. On the other hand, for the case of a strong friction, $\eta_f>1$, from the energy balance we have $\gamma\sim(\rho_fa_0)^{-1/2}<a_0$. As a result, for $\eta_f>1$ an impact of $\mathbf{f}_2$ on electron motion would be small in comparison with that produced by $\mathbf{f}_1$ and $\mathbf{f}_3$. However, we will keep the force $\mathbf{f}_2$ for completeness, which can also to some extent mimic the impact of $\mathbf{f}_s$ as we found that $\mathbf{f}_s$ and $\mathbf{f}_2$ could be comparable in our simulations.

Nonetheless, since $\rho_f\ll1$, an impact of radiation friction on electron trajectory is small unless we are dealing with the situation where electron trajectories are strongly diverging. Such strongly diverging electron trajectories are typical in the vicinity of electron trapping sites (attractors) considered in \cite{8JGKirk,10Agonoskov,11TZhEsirkepov,12SVBulanov2017}. Therefore, one could expect that for such case the impact of all the components of the radiation friction force on electron trajectory and, therefore, electron energy gain will be comparable.

To illustrate the impact of $\mathbf{f}_2$ and $\mathbf{f}_3$ components of the radiation friction force on the electron dynamics, we will consider the electromagnetic field in the form of a standing wave \cite{6SVBulanov,7ARBell,8JGKirk,9GLehmann,10Agonoskov,11TZhEsirkepov,12SVBulanov2017}, which is characterized by the following vector potential
\begin{equation}
\mathbf{A}=a_0cos(z)cos(t)\mathbf{e}_x,
\label{eq8}
\end{equation}
which will be employed to numerically solve Eqs.~(\ref{eq1}-\ref{eq4}) with and without $\mathbf{f}_2$ and $\mathbf{f}_3$ components. 

We notice that parameters $q$ and $\eta_f$ depend on the laser wave amplitude and wavenumber differently. To reveal an impact of $\mathbf{f}_2$ and $\mathbf{f}_3$ components more clearly for large friction force case, in our simulations we take $\lambda=1\mu m$ and $a_0=1000$ (corresponding to laser intensity of $I=1.37\times10^{24}Wcm^{-2}$) such that $\eta_f\approx12$ and $q$ in Eq.~(\ref{eq5}) will be of order unity and thus only modest variation of the function $Q(q)$ appears. As indicated by Eqs.~(\ref{eq1}-\ref{eq3}) the y-component of the electron equation of motion has only dissipated term originated from $\mathbf{f}_1$ for the vector potential in Eq.~(\ref{eq8}) and thus it is reasonable to set the y-component of electron momentum as zero which will be conserved and so is the y coordinate. Therefore, we need to only deal with 4-dimensional equations. Under such set up, electron dynamics with different initial coordinates of z (the x coordinate doesn't affect the electron motion and thus is chosen as zero initially) and momenta ($\mathbf{P}=P_x\mathbf{e}_x+P_z\mathbf{e}_z$) are investigated. 

\begin{figure}
\centering
\subfigure{
\label{fig1a}
\begin{minipage}[b]{0.22\textwidth}
\includegraphics[width=1\textwidth]{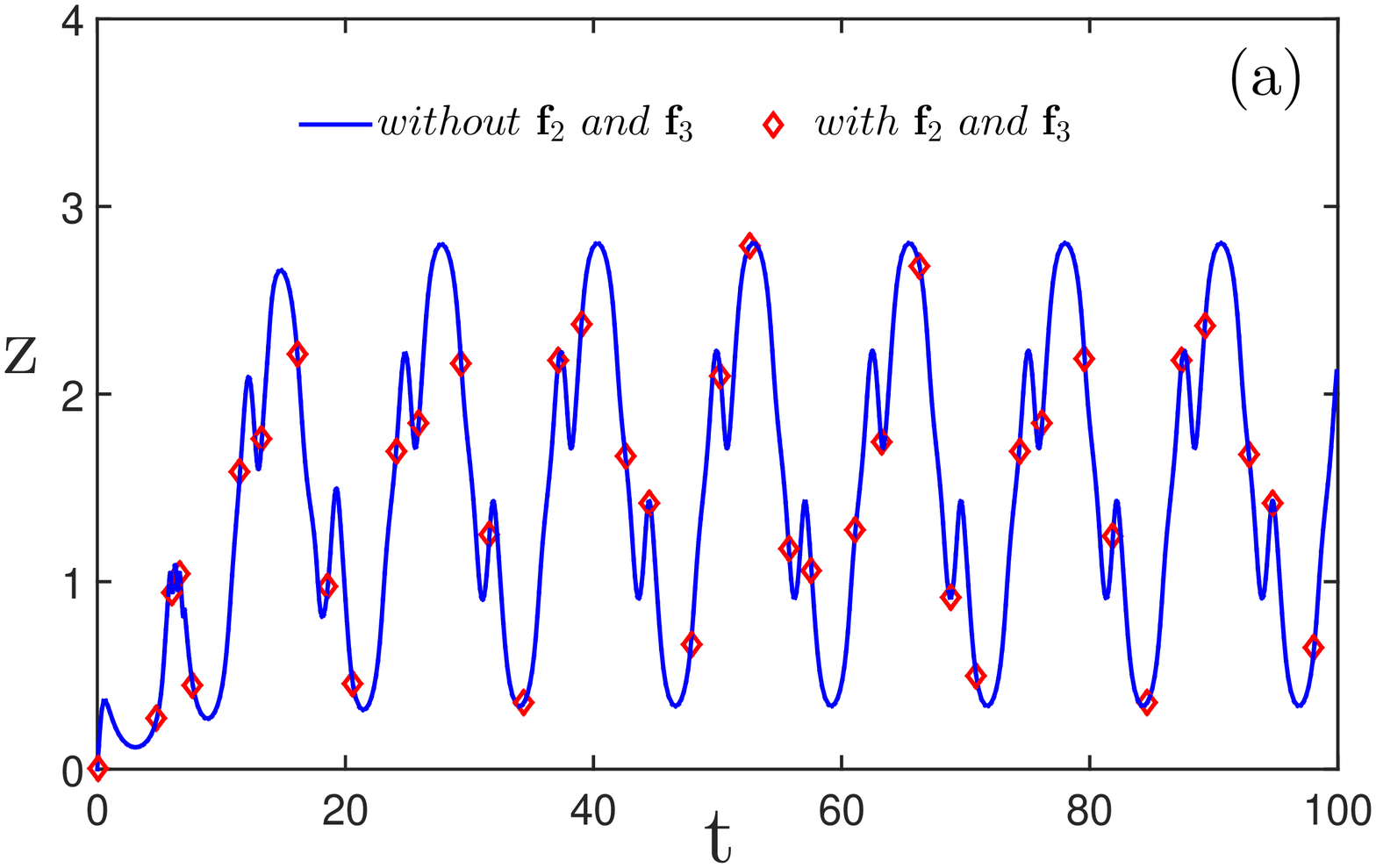}
\end{minipage}}
\subfigure{
\label{fig1b}
\begin{minipage}[b]{0.22\textwidth}
\includegraphics[width=1\textwidth]{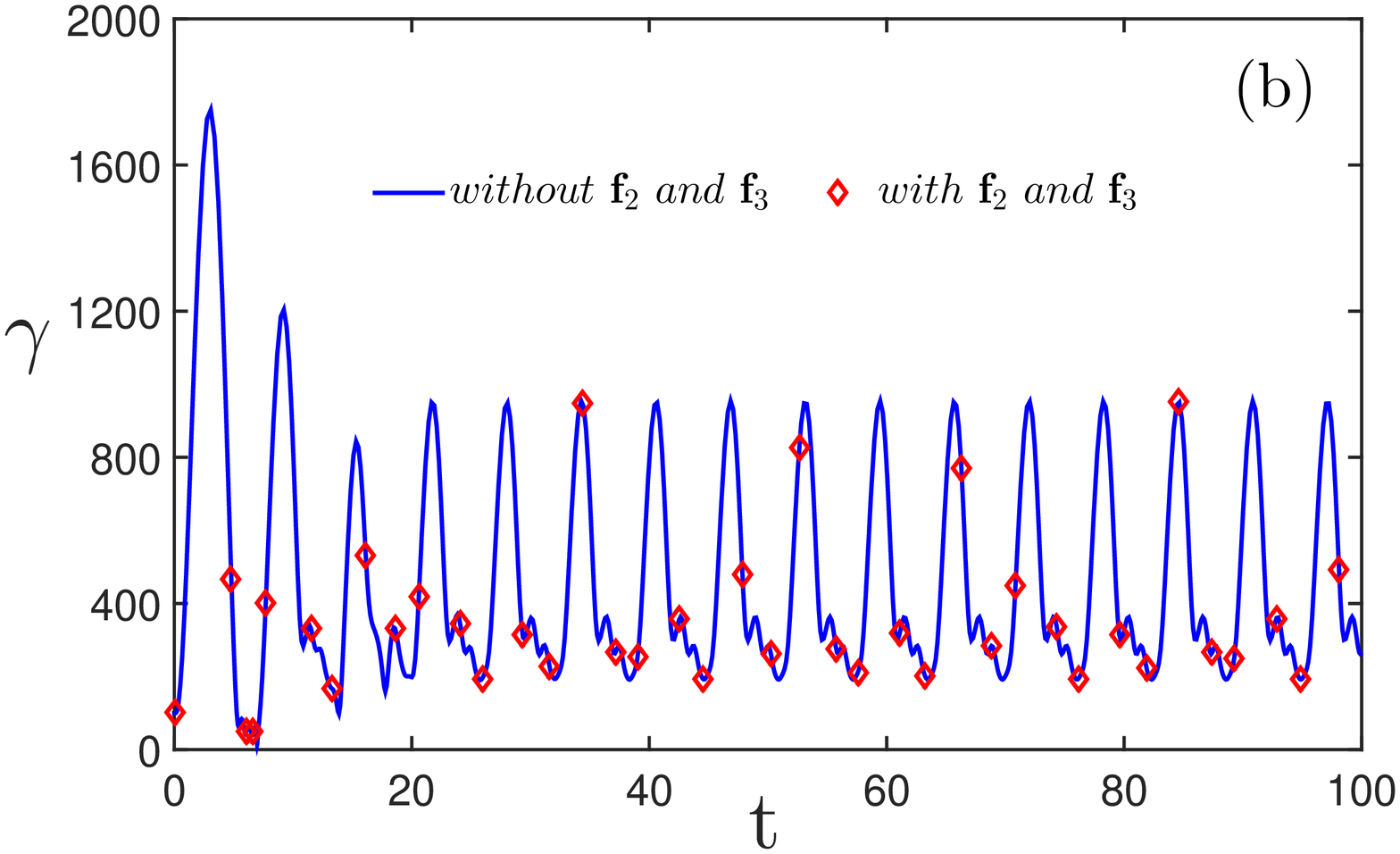}
\end{minipage}}
\caption{Electron dynamics for $\lambda=1\mu m$, $a_0=1000$ (laser intensity of $I=1.37\times10^{24}Wcm^{-2}$), and initial conditions $z(0)=0$, $P_x(0)=0$ and $P_z(0)=100$. The solid blue curve (red diamond marker) corresponds to the simulation results without (with) $\mathbf{f}_2$ and $\mathbf{f}_3$ components.}
\label{fig1}
\end{figure}

Shown in Fig.~\ref{fig1} is the results for electron with $z(0)=0$, $P_x(0)=0$ and $P_z(0)=100$, where electron performs periodic motion. The solid blue curve and red diamond marker denote, respectively, the results without and with $\mathbf{f}_2$ and $\mathbf{f}_3$ components. As we can see for this non-diverging electron trajectory, the impact of $\mathbf{f}_2$ and $\mathbf{f}_3$ components on such highly relativistic electron dynamics is completely negligible. 

\begin{figure}
\centering
\subfigure{
\label{fig2a}
\includegraphics[width=0.4\textwidth]{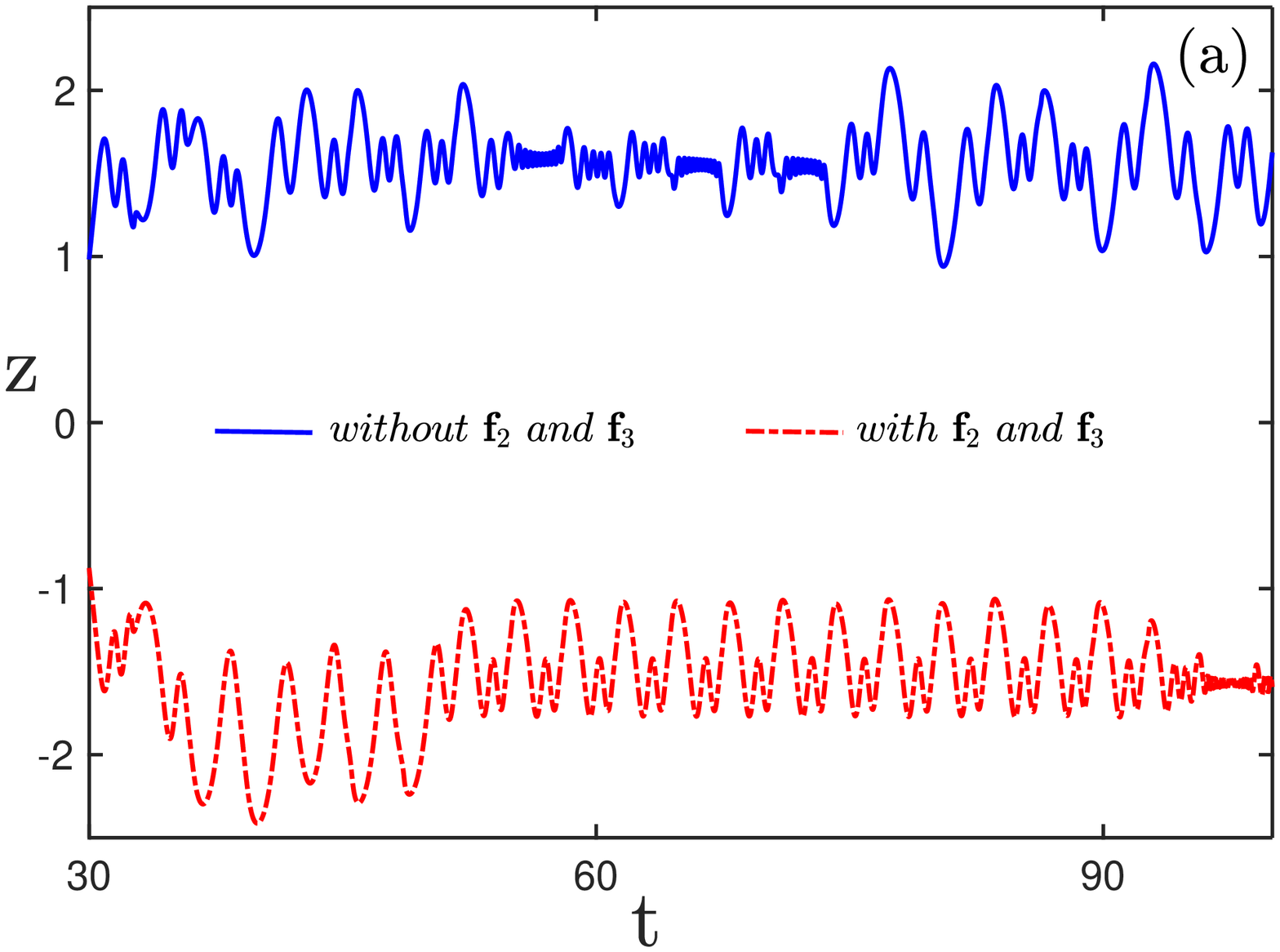}}
\subfigure{
\label{fig2b}
\includegraphics[width=0.4\textwidth]{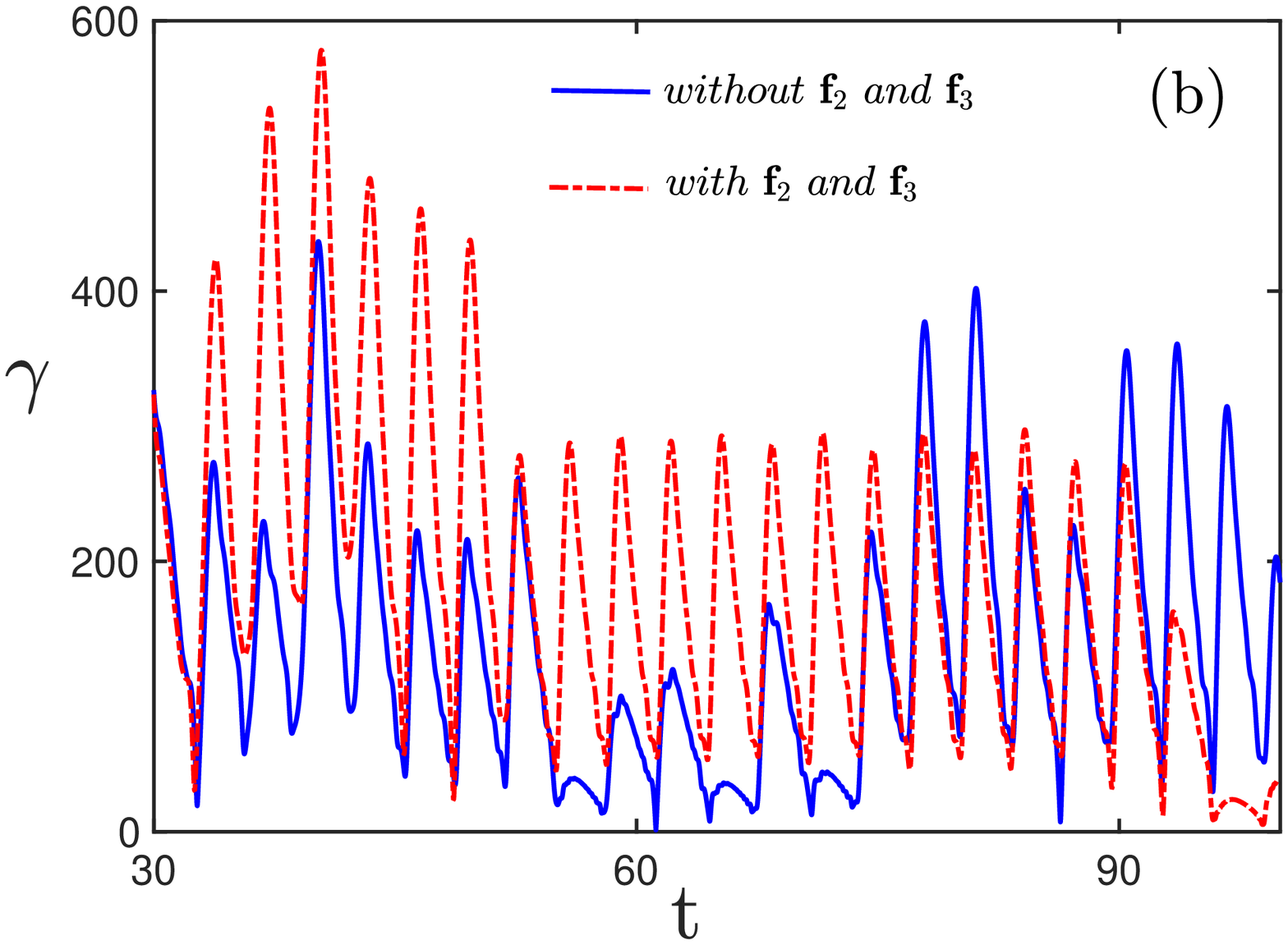}}
\subfigure{
\label{fig2c}
\includegraphics[width=0.4\textwidth]{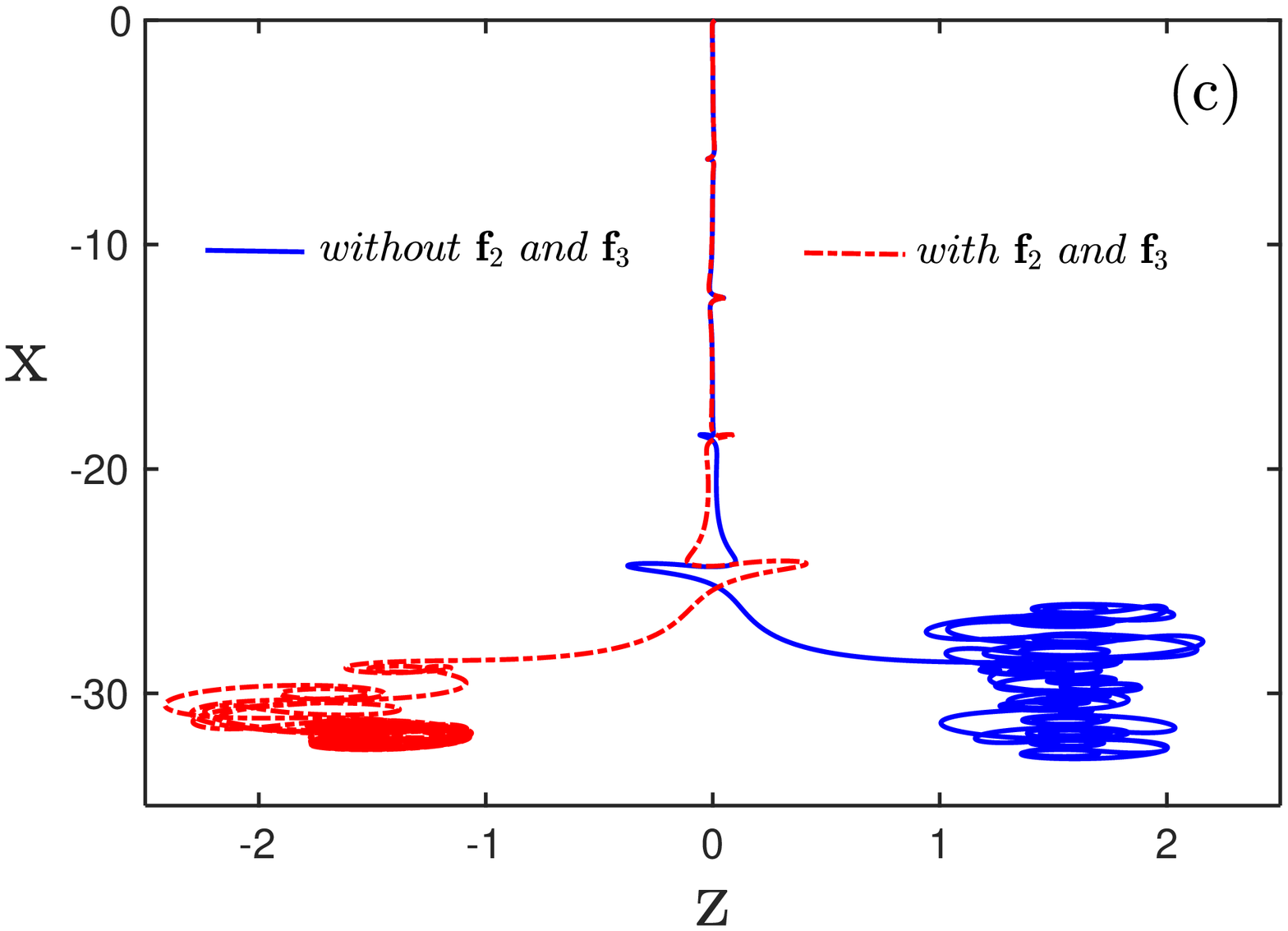}}
\caption{Electron dynamics for the same parameters with Fig.~\ref{fig1} but different initial conditions of $z(0)=0.01$, $P_x(0)=P_z(0)=0$. The solid blue and dash-dot red curves denote the simulation results without (with) $\mathbf{f}_2$ and $\mathbf{f}_3$ components, respectively.}
\label{fig2}
\end{figure}

However, this is not the case for electron with strongly diverging trajectories. For example, the electron motions for $z(0)=0.01$, and $P_x(0)=P_z(0)=0$ have been shown in Fig.~\ref{fig2} where the results without and with $\mathbf{f}_2$ and $\mathbf{f}_3$ components are displayed, respectively, by the solid blue and dash-dot red curves (we omit initial stage of electron motion for z coordinates and gamma-factors versus time where the gamma-factors are settling down). As one can see, $\mathbf{f}_2$ and $\mathbf{f}_3$ components of the radiation friction force make significant impact on all parameters shown in Fig.~\ref{fig2}. They adjust the electron trajectories [Fig.~\ref{fig2a}] and thus the energy gains [Fig.~\ref{fig2b}], where $\gamma<a_0$ in this region due to strong radiation friction effects. Fig.~\ref{fig2a} shows that, in accordance with the results of \cite{8JGKirk,10Agonoskov,11TZhEsirkepov,12SVBulanov2017}, due to an impact of radiation friction electron starts to be trapped in the vicinities of zero electric field at $z=\pi/2\pm n\pi$ with $n=0,1,2...$ (attractors). Whether including $\mathbf{f}_2$ and $\mathbf{f}_3$ components or not leads to electron ending up in different attractors (we found that the same attractor of $z=-\pi/2$ for the case with $\mathbf{f}_2$ and $\mathbf{f}_3$ components is obtained for electron motion with $\mathbf{f}_3$ but not $\mathbf{f}_2$). The divergence of electron trajectory under the impact of $\mathbf{f}_2$ and $\mathbf{f}_3$ is clearly seen in Fig.~\ref{fig2c} where electron trajectories are shown from $t=0$. However, such diverging effects depend on initial conditions (e.g., for initial coordinate $z(0)=0.02$ and the same momentum with Fig.~\ref{fig2} the electrons with and without $\mathbf{f}_2$ and $\mathbf{f}_3$ components fall into the same attractor near $z=\pi/2$).

\begin{figure}
\centering
\subfigure{
\label{fig3a}
\begin{minipage}[b]{0.4\textwidth}
\includegraphics[width=1\textwidth]{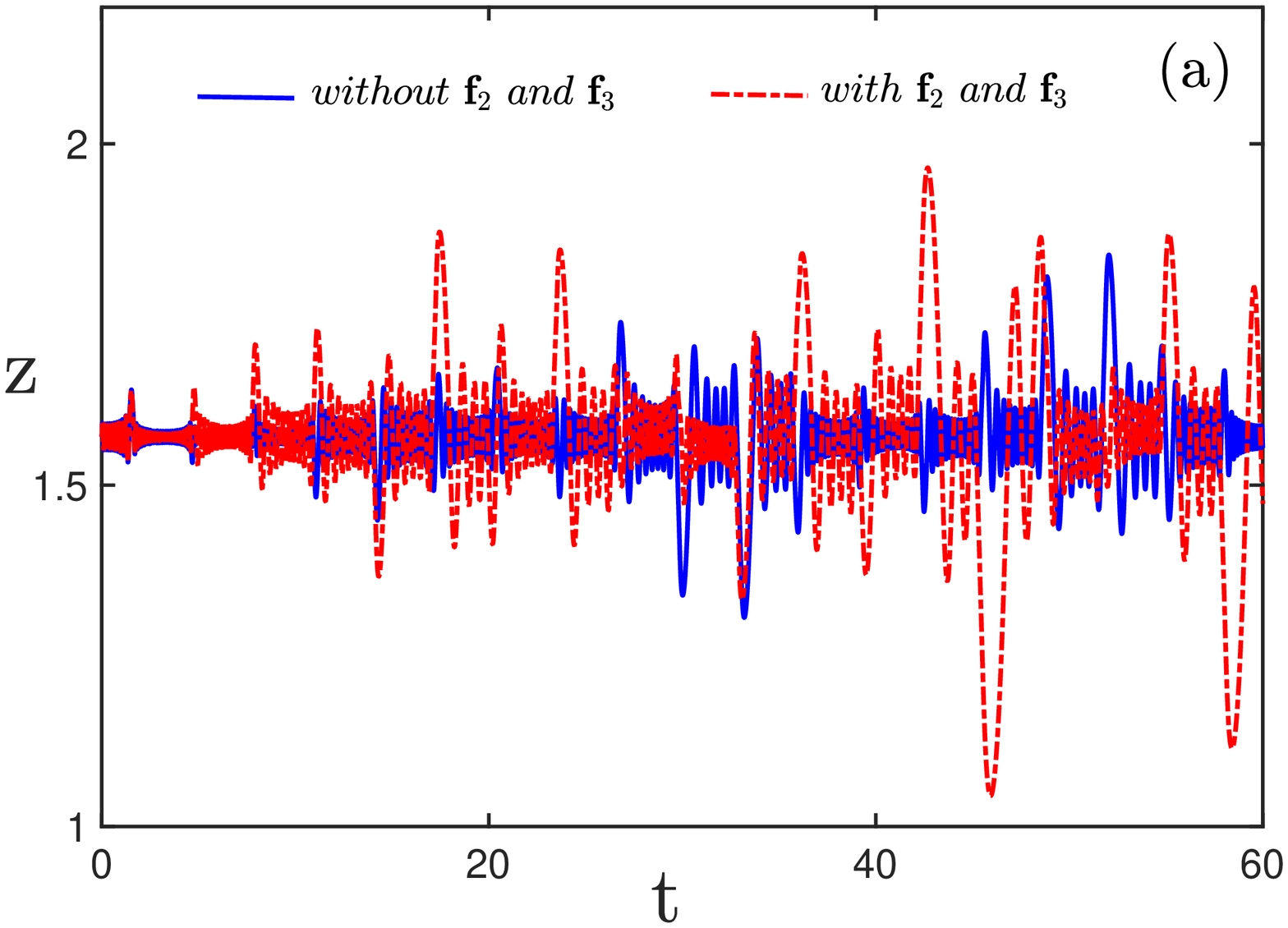} 
\end{minipage}}
\subfigure{
\label{fig3b}
\begin{minipage}[b]{0.4\textwidth}
\includegraphics[width=1\textwidth]{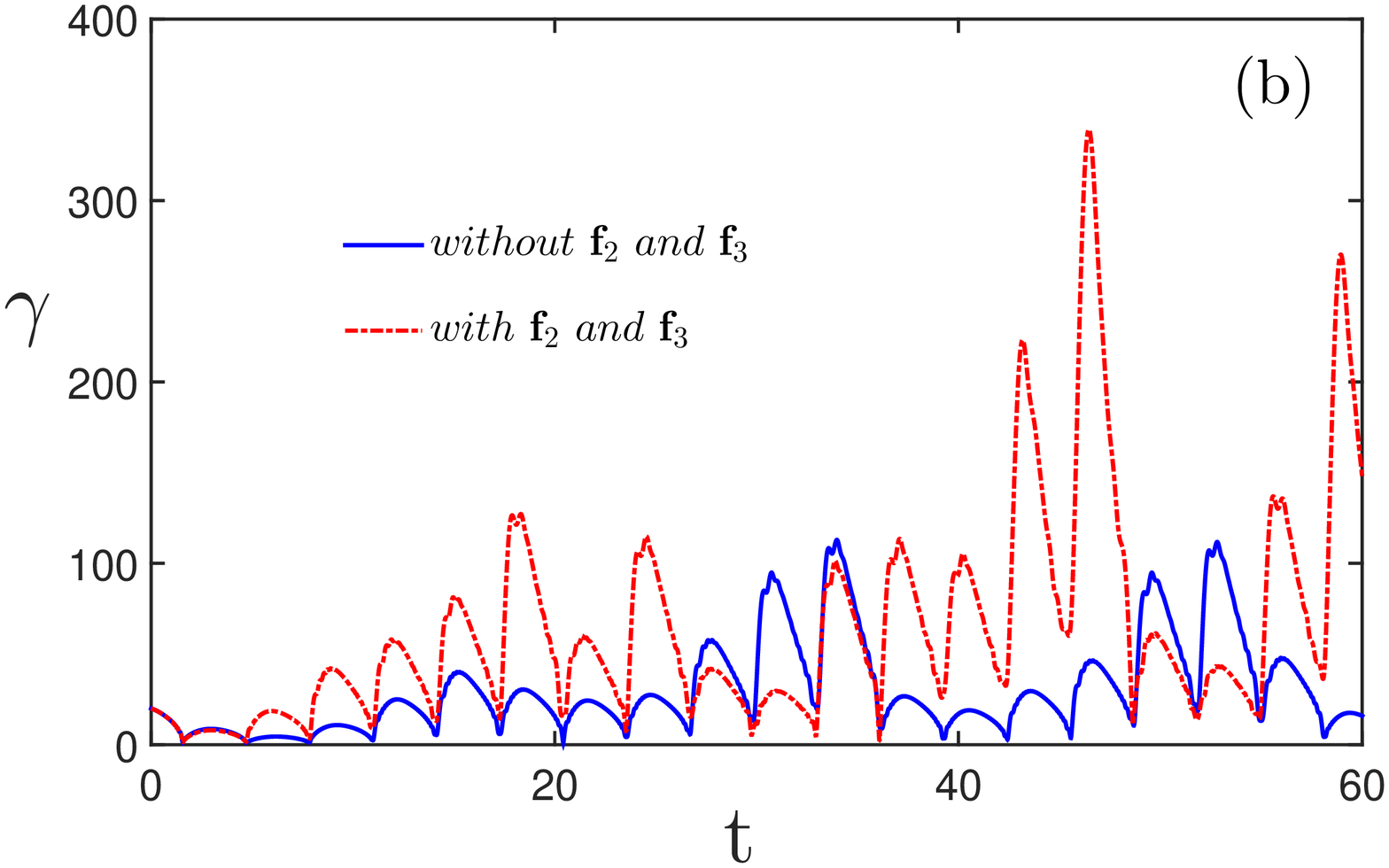}
\end{minipage}}
\caption{Electron dynamics for the same conditions with Fig.~\ref{fig2} but $z(0)=\pi/2$ and $P_z(0)=20$. }
\label{fig3}
\end{figure}

To examine the impact of $\mathbf{f}_2$ and $\mathbf{f}_3$ components on the electron dynamics in the same attractor, we initially put the electron into the attractor $z=\pi/2$ and take $P_x(0)=0$, $P_z(0)=20$. The results are shown in Fig.~\ref{fig3}, which demonstrates that $\mathbf{f}_2$ and $\mathbf{f}_3$ components could significantly affect the electron trajectory and thus energy gain even when they are around the same attractor. We notice that it is hard to obtain reliable long time simulation results of electron motion for such strongly diverging electron trajectories in Figs.~\ref{fig2} and \ref{fig3}, but the relatively short time simulations already exhibit the significant impacts of $\mathbf{f}_2$ and $\mathbf{f}_3$ components.

\begin{figure}[b]
\centering
\begin{minipage}{1\linewidth}
\includegraphics[width=1\textwidth]{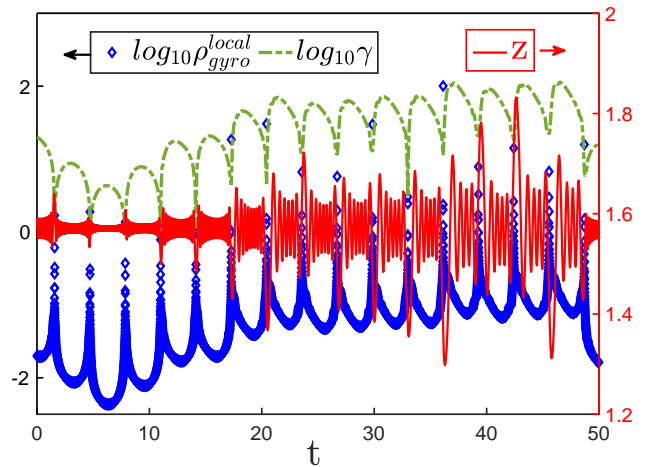}
\caption{The logarithms of electron gamma-factor (dashed green) and local gyro-radius $\rho_{gyro}^{local}$ (blue diamond) (quantified by the left y axis) and electron coordinate z (solid red) (corresponding to the right y axis) versus time for case of Fig.~\ref{fig2} without $\mathbf{f}_2$ and $\mathbf{f}_3$ components. }
\label{fig4}
\end{minipage}
\end{figure}

We see fast oscillations of electron trajectory at trapping sites [e.g., see Fig.~\ref{fig3a}], which were also observed in \cite{10Agonoskov,11TZhEsirkepov}. These oscillations are just the gyro-motion of relativistic electron in magnetic field in the vicinity of electric node as seen from Eq.~(\ref{eq7}). In our normalized units the gyro-frequency and gyro-radius can be expressed, respectively, as $\Omega_{gyro}=\left|B(t)\right|/\gamma$ and $\rho_{gyro}\approx\gamma/\left|B(t)\right|$, where the magnitude of the magnetic field directed in y-direction depends on time and z-coordinate as $B(t)=-a_0sin(z)cos(t)$ and the electron motion speed is approximately taken as unity. For the case of large radiation friction where $\gamma\ll a_0$ [e.g, see Fig.~\ref{fig3b}], for most of the period of electromagnetic wave we have $\Omega_{gyro}\gg 1$ and $\rho_{gyro}\ll 1$ in the vicinity of electric nodes. The adiabatic invariant $\mu\propto \left[\gamma^2v^2\right]/\left|B(t)\right|$ \cite{11LDLandaubook} is not conserved due to radiation friction, which continuously dissipates electron energy. However, for relatively short period of time around $B(t)=0$, electron gyro-radius becomes large ($\rho_{gyro}\ge 1$) and electron experiences large departure from trapping site [see Fig.~\ref{fig3a}], gaining kinetic energy from the laser [see Fig.~\ref{fig3b}]. More clearly one could see it from Fig.~\ref{fig4} where electron z-coordinate (solid red), gamma-factor (dash-dot green) and local gyro-radius $\rho_{gyro}^{local}$ (blue diamond) are displayed as the functions of time for the case of Fig.~\ref{fig3} without $\mathbf{f}_2$ and $\mathbf{f}_3$ components (notice that $\rho_{gyro}^{local}$ in Fig.~\ref{fig4} is not going to infinity when $B(t)=0$ because of discrete sampling). It shows that the electron begins to effectively exchange energy with laser when it performs large excursion away from the electric node and into the region with strong electric field.

In conclusion, we demonstrate that widely used assumption that for strong radiation friction $\eta_f>1$, the leading role in dynamics of highly relativistic electron is played by $\mathbf{f}_1$ component of the Landau-Lifshitz radiation friction force [see Eq.~(\ref{eq2})] is incorrect for the case of diverging electron trajectory. For latter case $\mathbf{f}_2$ and $\mathbf{f}_3$ components of the radiation friction force, having the same order as $\mathbf{f}_1$ component in equation of electron motion for trajectories (7), are equally important for electron both trajectory and energy gain. 

This work has been supported by the University of California Office of the President Lab Fee grant number LFR-17-449059.

\bibliography{references}

\begin{thebibliography}{16}%
\makeatletter
\providecommand \@ifxundefined [1]{%
 \@ifx{#1\undefined}
}%
\providecommand \@ifnum [1]{%
 \ifnum #1\expandafter \@firstoftwo
 \else \expandafter \@secondoftwo
 \fi
}%
\providecommand \@ifx [1]{%
 \ifx #1\expandafter \@firstoftwo
 \else \expandafter \@secondoftwo
 \fi
}%
\providecommand \natexlab [1]{#1}%
\providecommand \enquote  [1]{``#1''}%
\providecommand \bibnamefont  [1]{#1}%
\providecommand \bibfnamefont [1]{#1}%
\providecommand \citenamefont [1]{#1}%
\providecommand \href@noop [0]{\@secondoftwo}%
\providecommand \href [0]{\begingroup \@sanitize@url \@href}%
\providecommand \@href[1]{\@@startlink{#1}\@@href}%
\providecommand \@@href[1]{\endgroup#1\@@endlink}%
\providecommand \@sanitize@url [0]{\catcode `\\12\catcode `\$12\catcode
  `\&12\catcode `\#12\catcode `\^12\catcode `\_12\catcode `\%12\relax}%
\providecommand \@@startlink[1]{}%
\providecommand \@@endlink[0]{}%
\providecommand \url  [0]{\begingroup\@sanitize@url \@url }%
\providecommand \@url [1]{\endgroup\@href {#1}{\urlprefix }}%
\providecommand \urlprefix  [0]{URL }%
\providecommand \Eprint [0]{\href }%
\providecommand \doibase [0]{http://dx.doi.org/}%
\providecommand \selectlanguage [0]{\@gobble}%
\providecommand \bibinfo  [0]{\@secondoftwo}%
\providecommand \bibfield  [0]{\@secondoftwo}%
\providecommand \translation [1]{[#1]}%
\providecommand \BibitemOpen [0]{}%
\providecommand \bibitemStop [0]{}%
\providecommand \bibitemNoStop [0]{.\EOS\space}%
\providecommand \EOS [0]{\spacefactor3000\relax}%
\providecommand \BibitemShut  [1]{\csname bibitem#1\endcsname}%
\let\auto@bib@innerbib\@empty
\bibitem [{\citenamefont {Landau}\ and\ \citenamefont
  {Lifshitz}(2009)}]{11LDLandaubook}%
  \BibitemOpen
  \bibfield  {author} {\bibinfo {author} {\bibfnamefont {L.~D.}\ \bibnamefont
  {Landau}}\ and\ \bibinfo {author} {\bibfnamefont {E.~M.}\ \bibnamefont
  {Lifshitz}},\ }\href@noop {} {\emph {\bibinfo {title} {The Classical Theory
  of Fields, V. 2, Course of Theoretical Physics}}}\ (\bibinfo  {publisher}
  {Elsevier},\ \bibinfo {year} {2009})\BibitemShut {NoStop}%
\bibitem [{\citenamefont {Zel'dovich}(1975)}]{2YBZel}%
  \BibitemOpen
  \bibfield  {author} {\bibinfo {author} {\bibfnamefont {Y.~B.}\ \bibnamefont
  {Zel'dovich}},\ }\href@noop {} {\bibfield  {journal} {\bibinfo  {journal}
  {Sov. Phys. Uspekhi}\ }\textbf {\bibinfo {volume} {18}},\ \bibinfo {pages}
  {79} (\bibinfo {year} {1975})}\BibitemShut {NoStop}%
\bibitem [{\citenamefont {Sarachik}\ and\ \citenamefont
  {Schappert}(1970)}]{3ESarachikGSch}%
  \BibitemOpen
  \bibfield  {author} {\bibinfo {author} {\bibfnamefont {E.}~\bibnamefont
  {Sarachik}}\ and\ \bibinfo {author} {\bibfnamefont {G.}~\bibnamefont
  {Schappert}},\ }\href@noop {} {\bibfield  {journal} {\bibinfo  {journal}
  {Phys. Rev. D}\ }\textbf {\bibinfo {volume} {1}},\ \bibinfo {pages} {2738}
  (\bibinfo {year} {1970})}\BibitemShut {NoStop}%
\bibitem [{\citenamefont {Zhidkov}\ \emph {et~al.}(2002)\citenamefont
  {Zhidkov}, \citenamefont {Koga}, \citenamefont {Sasaki},\ and\ \citenamefont
  {Uesaka}}]{4Zhidkov}%
  \BibitemOpen
  \bibfield  {author} {\bibinfo {author} {\bibfnamefont {A.}~\bibnamefont
  {Zhidkov}}, \bibinfo {author} {\bibfnamefont {J.}~\bibnamefont {Koga}},
  \bibinfo {author} {\bibfnamefont {A.}~\bibnamefont {Sasaki}}, \ and\ \bibinfo
  {author} {\bibfnamefont {M.}~\bibnamefont {Uesaka}},\ }\href@noop {}
  {\bibfield  {journal} {\bibinfo  {journal} {Phys. Rev. Lett.}\ }\textbf
  {\bibinfo {volume} {88}},\ \bibinfo {pages} {185002} (\bibinfo {year}
  {2002})}\BibitemShut {NoStop}%
\bibitem [{\citenamefont {Bulanov}\ \emph {et~al.}(2011)\citenamefont
  {Bulanov}, \citenamefont {Esirkepov}, \citenamefont {Kando}, \citenamefont
  {Koga},\ and\ \citenamefont {Bulanov}}]{9bulanov2011lorentz}%
  \BibitemOpen
  \bibfield  {author} {\bibinfo {author} {\bibfnamefont {S.~V.}\ \bibnamefont
  {Bulanov}}, \bibinfo {author} {\bibfnamefont {T.~Z.}\ \bibnamefont
  {Esirkepov}}, \bibinfo {author} {\bibfnamefont {M.}~\bibnamefont {Kando}},
  \bibinfo {author} {\bibfnamefont {J.~K.}\ \bibnamefont {Koga}}, \ and\
  \bibinfo {author} {\bibfnamefont {S.~S.}\ \bibnamefont {Bulanov}},\
  }\href@noop {} {\bibfield  {journal} {\bibinfo  {journal} {Physical Review
  E}\ }\textbf {\bibinfo {volume} {84}},\ \bibinfo {pages} {056605} (\bibinfo
  {year} {2011})}\BibitemShut {NoStop}%
\bibitem [{\citenamefont {Bell}\ and\ \citenamefont {Kirk}(2008)}]{7ARBell}%
  \BibitemOpen
  \bibfield  {author} {\bibinfo {author} {\bibfnamefont {A.~R.}\ \bibnamefont
  {Bell}}\ and\ \bibinfo {author} {\bibfnamefont {J.~G.}\ \bibnamefont
  {Kirk}},\ }\href@noop {} {\bibfield  {journal} {\bibinfo  {journal} {Phys.
  Rev. Lett.}\ }\textbf {\bibinfo {volume} {101}},\ \bibinfo {pages} {200403}
  (\bibinfo {year} {2008})}\BibitemShut {NoStop}%
\bibitem [{\citenamefont {Lehmann}\ and\ \citenamefont
  {Spatschek}(2012)}]{9GLehmann}%
  \BibitemOpen
  \bibfield  {author} {\bibinfo {author} {\bibfnamefont {G.}~\bibnamefont
  {Lehmann}}\ and\ \bibinfo {author} {\bibfnamefont {K.~H.}\ \bibnamefont
  {Spatschek}},\ }\href@noop {} {\bibfield  {journal} {\bibinfo  {journal}
  {Phys. Rev. E.}\ }\textbf {\bibinfo {volume} {85}},\ \bibinfo {pages}
  {056412} (\bibinfo {year} {2012})}\BibitemShut {NoStop}%
\bibitem [{\citenamefont {Bulanov}\ \emph {et~al.}(2004)\citenamefont
  {Bulanov}, \citenamefont {Esirkepov}, \citenamefont {Koga},\ and\
  \citenamefont {Tajima}}]{6SVBulanov}%
  \BibitemOpen
  \bibfield  {author} {\bibinfo {author} {\bibfnamefont {S.~V.}\ \bibnamefont
  {Bulanov}}, \bibinfo {author} {\bibfnamefont {T.~Z.}\ \bibnamefont
  {Esirkepov}}, \bibinfo {author} {\bibfnamefont {J.}~\bibnamefont {Koga}}, \
  and\ \bibinfo {author} {\bibfnamefont {T.}~\bibnamefont {Tajima}},\
  }\href@noop {} {\bibfield  {journal} {\bibinfo  {journal} {Plasma Phys.
  Rep.}\ }\textbf {\bibinfo {volume} {30}},\ \bibinfo {pages} {196} (\bibinfo
  {year} {2004})}\BibitemShut {NoStop}%
\bibitem [{\citenamefont {Kirk}\ \emph {et~al.}(2009)\citenamefont {Kirk},
  \citenamefont {Bell},\ and\ \citenamefont {Arka}}]{8JGKirk}%
  \BibitemOpen
  \bibfield  {author} {\bibinfo {author} {\bibfnamefont {J.~G.}\ \bibnamefont
  {Kirk}}, \bibinfo {author} {\bibfnamefont {A.~R.}\ \bibnamefont {Bell}}, \
  and\ \bibinfo {author} {\bibfnamefont {I.}~\bibnamefont {Arka}},\ }\href@noop
  {} {\bibfield  {journal} {\bibinfo  {journal} {Plasma Phys. Contr. Fusion}\
  }\textbf {\bibinfo {volume} {51}},\ \bibinfo {pages} {085008} (\bibinfo
  {year} {2009})}\BibitemShut {NoStop}%
\bibitem [{\citenamefont {Gonoskov}\ \emph {et~al.}(2014)\citenamefont
  {Gonoskov}, \citenamefont {Bashinov}, \citenamefont {Gonoskov}, \citenamefont
  {Harvey}, \citenamefont {Ilderton}, \citenamefont {Kim}, \citenamefont
  {Marklund}, \citenamefont {Mourou},\ and\ \citenamefont
  {Sergeev}}]{10Agonoskov}%
  \BibitemOpen
  \bibfield  {author} {\bibinfo {author} {\bibfnamefont {A.}~\bibnamefont
  {Gonoskov}}, \bibinfo {author} {\bibfnamefont {A.}~\bibnamefont {Bashinov}},
  \bibinfo {author} {\bibfnamefont {I.}~\bibnamefont {Gonoskov}}, \bibinfo
  {author} {\bibfnamefont {C.}~\bibnamefont {Harvey}}, \bibinfo {author}
  {\bibfnamefont {A.}~\bibnamefont {Ilderton}}, \bibinfo {author}
  {\bibfnamefont {A.}~\bibnamefont {Kim}}, \bibinfo {author} {\bibfnamefont
  {M.}~\bibnamefont {Marklund}}, \bibinfo {author} {\bibfnamefont
  {G.}~\bibnamefont {Mourou}}, \ and\ \bibinfo {author} {\bibfnamefont
  {A.}~\bibnamefont {Sergeev}},\ }\href@noop {} {\bibfield  {journal} {\bibinfo
   {journal} {Phys. Rev. Lett.}\ }\textbf {\bibinfo {volume} {113}},\ \bibinfo
  {pages} {014801} (\bibinfo {year} {2014})}\BibitemShut {NoStop}%
\bibitem [{\citenamefont {Esirkepov}\ \emph {et~al.}(2015)\citenamefont
  {Esirkepov}, \citenamefont {Bulanov}, \citenamefont {Koga}, \citenamefont
  {Kando}, \citenamefont {Kondo}, \citenamefont {Rosanov}, \citenamefont
  {Korn},\ and\ \citenamefont {Bulanov}}]{11TZhEsirkepov}%
  \BibitemOpen
  \bibfield  {author} {\bibinfo {author} {\bibfnamefont {T.~Z.}\ \bibnamefont
  {Esirkepov}}, \bibinfo {author} {\bibfnamefont {S.~S.}\ \bibnamefont
  {Bulanov}}, \bibinfo {author} {\bibfnamefont {J.~K.}\ \bibnamefont {Koga}},
  \bibinfo {author} {\bibfnamefont {M.}~\bibnamefont {Kando}}, \bibinfo
  {author} {\bibfnamefont {K.}~\bibnamefont {Kondo}}, \bibinfo {author}
  {\bibfnamefont {N.~N.}\ \bibnamefont {Rosanov}}, \bibinfo {author}
  {\bibfnamefont {G.}~\bibnamefont {Korn}}, \ and\ \bibinfo {author}
  {\bibfnamefont {S.~V.}\ \bibnamefont {Bulanov}},\ }\href@noop {} {\bibfield
  {journal} {\bibinfo  {journal} {Phys. Lett. A}\ }\textbf {\bibinfo {volume}
  {379}},\ \bibinfo {pages} {2044} (\bibinfo {year} {2015})}\BibitemShut
  {NoStop}%
\bibitem [{\citenamefont {Bulanov}\ \emph {et~al.}(2017)\citenamefont
  {Bulanov}, \citenamefont {Esirkepov}, \citenamefont {Koga}, \citenamefont
  {Bulanov}, \citenamefont {Gong}, \citenamefont {Yan},\ and\ \citenamefont
  {Kando}}]{12SVBulanov2017}%
  \BibitemOpen
  \bibfield  {author} {\bibinfo {author} {\bibfnamefont {S.~V.}\ \bibnamefont
  {Bulanov}}, \bibinfo {author} {\bibfnamefont {T.~Z.}\ \bibnamefont
  {Esirkepov}}, \bibinfo {author} {\bibfnamefont {J.~K.}\ \bibnamefont {Koga}},
  \bibinfo {author} {\bibfnamefont {S.~S.}\ \bibnamefont {Bulanov}}, \bibinfo
  {author} {\bibfnamefont {Z.}~\bibnamefont {Gong}}, \bibinfo {author}
  {\bibfnamefont {X.~Q.}\ \bibnamefont {Yan}}, \ and\ \bibinfo {author}
  {\bibfnamefont {M.}~\bibnamefont {Kando}},\ }\href@noop {} {\bibfield
  {journal} {\bibinfo  {journal} {J. Plasma Phys.}\ }\textbf {\bibinfo {volume}
  {83}},\ \bibinfo {pages} {905830202} (\bibinfo {year} {2017})}\BibitemShut
  {NoStop}%
\bibitem [{\citenamefont {Ji}\ \emph {et~al.}(2014)\citenamefont {Ji},
  \citenamefont {Pukhov}, \citenamefont {Kostyukov}, \citenamefont {Shen},\
  and\ \citenamefont {Akli}}]{5LLjiAPukov}%
  \BibitemOpen
  \bibfield  {author} {\bibinfo {author} {\bibfnamefont {L.~L.}\ \bibnamefont
  {Ji}}, \bibinfo {author} {\bibfnamefont {A.}~\bibnamefont {Pukhov}}, \bibinfo
  {author} {\bibfnamefont {I.~Y.}\ \bibnamefont {Kostyukov}}, \bibinfo {author}
  {\bibfnamefont {B.~F.}\ \bibnamefont {Shen}}, \ and\ \bibinfo {author}
  {\bibfnamefont {K.}~\bibnamefont {Akli}},\ }\href@noop {} {\bibfield
  {journal} {\bibinfo  {journal} {Phys. Rev. Lett.}\ }\textbf {\bibinfo
  {volume} {112}},\ \bibinfo {pages} {145003} (\bibinfo {year}
  {2014})}\BibitemShut {NoStop}%
\bibitem [{\citenamefont {Tamburini}\ \emph {et~al.}(2010)\citenamefont
  {Tamburini}, \citenamefont {Pegoraro}, \citenamefont {Piazza}, \citenamefont
  {Keitel},\ and\ \citenamefont {Macchi}}]{14MTamburini}%
  \BibitemOpen
  \bibfield  {author} {\bibinfo {author} {\bibfnamefont {M.}~\bibnamefont
  {Tamburini}}, \bibinfo {author} {\bibfnamefont {F.}~\bibnamefont {Pegoraro}},
  \bibinfo {author} {\bibfnamefont {A.~D.}\ \bibnamefont {Piazza}}, \bibinfo
  {author} {\bibfnamefont {C.~H.}\ \bibnamefont {Keitel}}, \ and\ \bibinfo
  {author} {\bibfnamefont {A.}~\bibnamefont {Macchi}},\ }\href@noop {}
  {\bibfield  {journal} {\bibinfo  {journal} {New J. Phys.}\ }\textbf {\bibinfo
  {volume} {12}},\ \bibinfo {pages} {123005} (\bibinfo {year}
  {2010})}\BibitemShut {NoStop}%
\bibitem [{\citenamefont {Mahajan}\ \emph {et~al.}(2015)\citenamefont
  {Mahajan}, \citenamefont {Asenjo},\ and\ \citenamefont
  {Hazeltine}}]{15SMMahajan}%
  \BibitemOpen
  \bibfield  {author} {\bibinfo {author} {\bibfnamefont {S.~M.}\ \bibnamefont
  {Mahajan}}, \bibinfo {author} {\bibfnamefont {F.~A.}\ \bibnamefont {Asenjo}},
  \ and\ \bibinfo {author} {\bibfnamefont {R.~D.}\ \bibnamefont {Hazeltine}},\
  }\href@noop {} {\bibfield  {journal} {\bibinfo  {journal} {Mon. Not. R.
  Astron. Soc.}\ }\textbf {\bibinfo {volume} {446}},\ \bibinfo {pages} {4112}
  (\bibinfo {year} {2015})}\BibitemShut {NoStop}%
\bibitem [{\citenamefont {Wen}\ \emph {et~al.}(2016)\citenamefont {Wen},
  \citenamefont {Bauke},\ and\ \citenamefont {Keite}}]{16MWenHBauke}%
  \BibitemOpen
  \bibfield  {author} {\bibinfo {author} {\bibfnamefont {M.}~\bibnamefont
  {Wen}}, \bibinfo {author} {\bibfnamefont {H.}~\bibnamefont {Bauke}}, \ and\
  \bibinfo {author} {\bibfnamefont {C.~H.}\ \bibnamefont {Keite}},\ }\href@noop
  {} {\bibfield  {journal} {\bibinfo  {journal} {Sci. Rep.}\ }\textbf {\bibinfo
  {volume} {6}},\ \bibinfo {pages} {31624} (\bibinfo {year}
  {2016})}\BibitemShut {NoStop}%
\end{thebibliography}%

\end{document}